\documentclass[sigconf]{acmart}

\AtBeginDocument{%
  }

\copyrightyear{2025}
\acmYear{2025}
\setcopyright{cc}
\setcctype{by}
\acmConference[SIGIR '25]{Proceedings of the 48th International ACM SIGIR Conference on Research and Development in Information Retrieval}{July 13--18, 2025}{Padua, Italy}
\acmBooktitle{Proceedings of the 48th International ACM SIGIR Conference on Research and Development in Information Retrieval (SIGIR '25), July 13--18, 2025, Padua, Italy}
\acmDOI{10.1145/3726302.3729966}
\acmISBN{979-8-4007-1592-1/2025/07}

\usepackage[inline]{enumitem}
\usepackage{pbox}
\usepackage{balance}
\newlist{inlinelist}{enumerate*}{1}
\setlist*[inlinelist,1]{label=\roman*),itemjoin={{, }},itemjoin*={{, and }}}
\usepackage{url}
\usepackage{multirow}
\usepackage{multicol}
\usepackage{booktabs}
\usepackage{tikz}
\usepackage{xcolor}
\usepackage{lipsum}
\usepackage{acronym}
\usepackage{subcaption}
\usepackage[inline]{enumitem}
\usepackage{xspace}
\usepackage{arydshln}
\usepackage{array}

\usepackage{newfloat}
\DeclareFloatingEnvironment[fileext=lop, listname={List of prompts}, 
                            name=Prompt, placement=h]{prompt}

\usepackage{listings}
\usepackage{xcolor}
\newcommand{\header}[1]{\vspace{2mm}\noindent\textbf{#1}}

\acrodef{CS}{Conversational Search}
\acrodef{CSA}{Conversational Search Agent}
\acrodef{PTKB}{Personal Text Knowledge Base}
\acrodef{TREC}{TExt Retrieval Conference}
\acrodef{iKAT}{Interactive Knowledge Assistance Track}
\acrodef{CAsT}{Conversational Assistance Track}
\acrodef{NIST}{National Institute of Standards and Technology}
\acrodef{LLM}{Large Language Model}
\acrodef{LSR}{Learned Sparse Retrieval}

\acrodef{IR}{Information Retrieval}
\acrodef{NLP}{Natural Language Processing}
\acrodef{PEFT}{Parameter-Efficient Fine-Tuning}
\acrodef{ICL}{In-Context Learning}
\acrodef{LoRA}{Low-Rank Adaptation}

\acrodef{CQR}{Conversational Query Rewriting}

\acrodef{MSE}{Mean Square Error}

\newcommand{\spladis}{DiSCo\xspace}
\newcommand{\qrecc}{QReCC\xspace}
\newcommand{\topiocqa}{TopiOCQA\xspace}

\newcolumntype{P}[1]{>{\centering\arraybackslash}p{#1}}

\begin{document}

\title{DiSCo: LLM Knowledge Distillation for Efficient Sparse Retrieval in Conversational Search}

\author{Simon Lupart}
\orcid{0009-0008-2383-4557}
\affiliation{%
  \institution{University of Amsterdam}
  \city{Amsterdam}
  \country{Netherlands}}
\email{s.c.lupart@uva.nl}

\author{Mohammad Aliannejadi}
\orcid{0000-0002-9447-4172}
\affiliation{%
  \institution{University of Amsterdam}
  \city{Amsterdam}
  \country{Netherlands}}
\email{m.aliannejadi@uva.nl}

\author{Evangelos Kanoulas}
\orcid{0000-0002-8312-0694}
\affiliation{%
  \institution{University of Amsterdam}
  \city{Amsterdam}
  \country{Netherlands}}
\email{e.kanoulas@uva.nl}

\renewcommand{\shortauthors}{Simon Lupart, Mohammad Aliannejadi, and Evangelos Kanoulas}

\begin{abstract}

Conversational Search (CS) involves retrieving relevant documents from a corpus while considering the conversational context, integrating retrieval with context modeling. Recent advancements in Large Language Models (LLMs) have significantly enhanced CS by enabling query rewriting based on conversational context. However, employing LLMs during inference poses efficiency challenges. Existing solutions mitigate this issue by distilling embeddings derived from human-rewritten queries, focusing primarily on learning the context modeling task. These methods, however, often separate the contrastive retrieval task from the distillation process, treating it as an independent loss term. To overcome these limitations, we introduce DiSCo (Distillation of Sparse Conversational retrieval), a novel approach that unifies retrieval and context modeling through a relaxed distillation objective. Instead of relying exclusively on representation learning, our method distills similarity scores between conversations and documents, providing more freedom in the representation space and better leveraging the contrastive nature of document relevance. Extensive experiments on Learned Sparse Retrieval (LSR) across five CS datasets demonstrate that DiSCo achieves substantial improvements in both in-domain and out-of-domain retrieval tasks, achieving up to a six-point gain in recall for out-of-domain datasets over state-of-the-art methods. Additionally, DiSCo employs a multi-teacher distillation strategy, using multiple LLMs as teachers, further enhancing performance and surpassing the individual teachers in in-domain settings. Furthermore, analysis of model sparsity reveals that DiSCo allows for more effective control over the sparsity of the trained models.

\end{abstract}



\maketitle

\begin{figure}[t] 
    \centering
    \begin{tikzpicture}
        \definecolor{darkgreen}{rgb}{0.0, 0.5, 0.0}
        \definecolor{darkred}{rgb}{0.8, 0.1, 0.1}
        \definecolor{darkblue}{rgb}{0.12, 0.56, 1.0}
        \draw[very thin, gray] (0,0) grid (4.9,4.5);
        \node at (0,-0.3) [above left] {$o$};
        \fill (1,1) circle (2pt);
        \draw[-][line width=0.5mm] (0,0) -- (2,2);
        \node at (1,1) [below right] {$\mathbf{E_{d}(d)}$};

        \draw[domain=0.5:3.7, samples=2, color=darkblue, line width=0.5mm] plot (\x, {4-\x});
        \draw[->][>=latex, color=darkred, line width=0.5mm] (3.3,2.5) -- (3.5,0.5);
        \fill (3.5,0.5) circle (2pt);
        \node at (3.7,0.5) [right] {$\mathbf{E_{q}(q_{rw})}$};
        \draw[->][>=latex, color=darkgreen, line width=0.5mm,dashed] (3.3,2.5) -- (1.5,2.5);

        \draw[->][>=latex, color=darkgreen, line width=0.5mm,dashed] (3.3,2.5) -- (3,1);

        \draw[->][>=latex, color=darkgreen, line width=0.5mm,dashed] (3.3,2.5) -- (2.3,1.7);
        \fill (3.3,2.5) circle (2pt);
        \node at (3.3,2.4) [above right] {$\mathbf{\tilde{E}_{q}(q_{conv})}$};
        
        \node[] at (0.4,3.4) [above right] {$\mathbf{Y=\{ X \in \mathbb{R}^2 \mid X^\top E_d(d) = s_{q_{rw}} \}}$};
        
        \draw[->] (0,0) -- (0.5,0) node[below right] {$x$};
        \draw[->] (0,0) -- (0,0.5) node[above left] {$y$};

        \draw[->][>=latex, color=darkred, line width=0.5mm] (-1,5) -- (-0.25,5);
        \node at (-0.25,5) [right] {convDR};
        
        \draw[->][>=latex, color=darkgreen, line width=0.5mm,dashed] (1.25,5) -- (2,5);
        \node at (2,5) [right] {DiSCo};
        
        \draw[-][>=latex, color=darkblue, line width=0.6mm] (3.25,5) -- (4,5);
        \node at (4,5) [right] {Hyperplane};

        \coordinate (P1) at (2, 2); 
        \coordinate (P2) at (1.85, 1.85); 
        \coordinate (P3) at (2.15,1.85);
        \coordinate (P4) at (2,1.7);
        \draw[thick] (P2) -- (P4) -- (P3);

        
        
        


    \end{tikzpicture}
    \caption{Similarity Score Distillation in $\mathbb{R}^2$. Existing loss functions bound representation of the full conversation representation to converge to a single rewrite representation (convDR, red arrow), while if we consider document $\mathbf{d}$ as anchor, an infinite number of representations, other than $\mathbf{E_{q}(q_{rw})}$, have the same similarity with $\mathbf{d}$ ($\mathbf{Y}$, blue hyperplane). \spladis allows the model to converge to the best representation from the $\mathbf{Y}$ hyperplane (green arrows), as a relaxation.}
    
    \label{fig:distil}
\end{figure}

\section{Introduction}

\ac{CS} is a well-established task, that has seen major improvements recently, thanks to the development of \acp{LLM}~\cite{mo2024chiqcontextualhistoryenhancement,mao2023largelanguagemodelsknowllm4cs,10.1145/3626772.3657860}. The goal of the \ac{CS} task is to retrieve relevant documents from a corpus within a conversational context, in response to the user's latest utterance. While sharing similarities with ad-hoc retrieval, the main challenge of \ac{CS} remains to model the conversational context~\cite{10.1145/3477495.3532678/convIR,10.1145/3020165.3020183,ye-etal-2023-enhancing}.
More specifically, as the conversation advances, the context becomes longer and noisier, often with topic switches and language ambiguities, making the last user utterance complex to resolve and understand for retrieval systems~\cite{DBLP:journals/corr/abs-2201-05176/neuralcs,adlakha-etal-2022-topiocqa,10.1145/3397271.3401206/cast19}.

Training retrieval models for this task is, however, challenging, because of the lack of large-scale conversational datasets, making human annotation an essential component of the process.
In particular, a lot of effort has been invested first to create conversations with passage relevance judgment~\cite{10.1145/3397271.3401206/cast19,10.1145/3626772.3657860}, and then to rewrite for each user utterance the contextualized version of each query (rewritten query) as the optimal denoised query of each turn \cite{elgohary-etal-2019-unpack}. Datasets with rewrites enable us to learn the \ac{CQR} task by auto-regressive models (e.g., T5 \cite{DBLP:journals/corr/abs-1910-10683/T5}), and then only pass the rewritten queries to retrieval models instead of the full noisy conversations. However, this two-step approach -- rewrite and retrieve -- is not efficient~\cite{10.1145/3404835.3462856}, and may lead to information loss and error within the rewrite phase that can propagate to the retrieval phase. Hence, the need for unified retrieval models that do both tasks together in the representation space~\cite{penha2020challenges,mo2024aligningqueryrepresentationrewritten}.

Approaches such as ConvDR \cite{10.1145/3404835.3462856}, coSPLADE \cite{hai2024cospladecontextualizingspladeconversational} and LeCoRe \cite{10.1145/3543507.3583265} all learn both conversational context modeling and retrieval tasks within the representation space, either in a dense or sparse embedding space. As illustrated in Figure~\ref{fig:distil}, they use a distillation objective enforcing the conversation representations to converge to the representations of gold human-rewritten queries (red arrow in the figure).
This objective is, however, restrictive and assumes gold-rewritten query representations are the \textit{only} optimal rewriting and \textit{only} learning target in the representations space, leaving little freedom for the model to further learn and optimize the representations. \spladis, differently from ConvDR relaxes the distillation targeting an entire hyperplane (blue hyperplane in the figure) instead of a single representation. This is achieved by focusing on query document similarity scores, rather than solely on the rewrite. Besides, previous work \cite{10.1145/3404835.3462856,hai2024cospladecontextualizingspladeconversational,10.1145/3543507.3583265} distills query rewrite independently from the ranking objective. We thus propose to fill this gap, with a single distillation loss that unifies both context modeling and ranking tasks, while not limiting the model to learn one single representation.

Our method learns to distill the similarity scores between rewritten queries and documents rather than query representations directly. By distilling similarities, we first align with the contrastive nature of the ranking task~\cite{10.1145/1277741.1277794/proxy,DBLP:journals/corr/abs-2102-03732/contrastive,wang2020understanding}, but we also relax the training objective toward the final goal of retrieval, which is to compute similarities between queries and documents~\cite{10.5555/1577069.1577078,Koch2015SiameseNN,DBLP:journals/corr/SongXJS15}. This relaxation allows the student to further learn the context modeling task, considering relevant and irrelevant documents from the corpus, rather than mapping all representations to the human representations. This objective also follows the precepts from \citet{hofstatter2020improving} on distillation for ad-hoc retrieval and can benefit from hard negatives within the distillation loss.
Both dense and sparse methods could be subject to this relaxation; however, in this work, we focus on learned sparse architectures~\cite{snrm}, such as SPLADE~\cite{formal2021splade,lassance2024spladev3}, as the relaxation would have more degrees of freedom due to the high dimensionality of the representations for LSR.

Besides, the proposed relaxation also reduces the level of constraints on the teachers, as any method producing similarity scores could be used for the distillation. In the context of \ac{CS}, this allows the model to learn from multiple \ac{LLM} rewrites, by fusing similarity scores of several teachers into a single score to be distilled. To the best of our knowledge, our work is the first to distill the knowledge of multiple LLM teachers for query rewriting.
This also distinguishes our method from the work from~\citet{hofstatter2020improving}, as we distill from \ac{LLM} knowledge through the rewrites.

Through our work, we make the following contributions: 
\begin{itemize} 
    \item We propose \spladis, a relaxation of the training objective of \ac{CS} model distilling similarities of rewritten queries rather than representations.~\footnote{Models checkpoints and code at \url{https://github.com/SimonLupart/disco-conv-splade}}
    \item We propose to distill knowledge from multiple teachers, unlocking the potential of mixtures of LLMs in \ac{CS}.
    \item We evaluate the effectiveness of \spladis on in-domain (QReCC, TopiOCQA) and out-of-domain (CAsT 2020, 2022, and iKAT 2023) datasets, achieving state-of-the-art performance.
    \item We analyze the sparsity of the learned representations compared to those of the original teacher models, and demonstrate the efficiency gains of the approach.
\end{itemize}

Our results demonstrate that our \textbf{Di}stillation of \textbf{S}parse \textbf{Co}n-versational retrieval, \textbf{DiSCo}, leads to in-domain performance improvements, through our relaxation of the distillation loss. 
We also see improved generalization capacities with 12\% gains on recall and 16\% on precision compared to previous zero-shot models (LeCoRe and QRACDR) on CAsT 2020.
We demonstrate that distilling from multiple LLM teachers for the query rewriting task brings further performance gains, compared to single LLM teacher.
Finally, thanks to the increased freedom on the representations, we show that we can better control the model sparsity, even for long contexts, in deeper turns of the conversations.

\section{Related Work}
\header{Conversational Search (CS)} differentiates itself from ad-hoc search primarily through the nature of the input. Conversational history can grow very long with turns' dependencies, whereas search engine queries are typically concise, limited to a few words~\cite{10.1145/3406522.3446027,lupart2023msshiftanalysismsmarco}. The main challenge in addressing \ac{CS} is modeling the conversational context, to only represent useful information from the previous turns~\cite{Wu2021CONQRRCQ}. Two possibilities exist to learn this noise reduction: first on the token level, with generative models that learn to generate contextualized queries from past conversations~\cite{elgohary-etal-2019-unpack}, or within the representation spaces~\cite{10.1145/3404835.3462856} of retrieval models~\cite{surveyCS}.

\header{Query rewriting.} Learning to resolve the conversational history is difficult, as no or very few conversational search engines are in production today, limiting the availability of user-interaction data.
The effort is thus mostly based on human-annotated data. One of the first conversational datasets are
QuAC~\cite{DBLP:journals/corr/abs-1808-07036/quac} and ORConvQA~\cite{10.1145/3397271.3401110/orconvqa}, where human annotators were tasked to create conversations out of existing documents, assuming a human-machine conversation over the topic of the document. Each paragraph could potentially be an utterance and relevant passages would be inferred by construction. Later CANARD~\cite{elgohary-etal-2019-unpack} was released on top of ORConvQA, consisting of human rewrites of each utterance of the conversations, to help learn the context modeling task. This defined one of the main sub-tasks of CS: Conversational Query Rewriting (CQR)~\cite{elgohary-etal-2019-unpack,ye-etal-2023-enhancing}. Following studies learned the \ac{CQR} task at the token level -- generating automatic query rewrites based on the conversation history -- on autoregressive language models (e.g., T5 or BART)~\cite{mao-etal-2023-search,Wu2021CONQRRCQ}. Similar methods~\cite{mo-etal-2023-convgqr} used language models to answer the query, before resolving the context. Today, work is still being done to solve the CQR task, using \acp{LLM} in zero- or few-shot fashion. CHIQ~\cite{mo2024chiqcontextualhistoryenhancement} proposes to decompose the context modeling task into several simpler sub-tasks for the \acp{LLM} -- history enhancement, answer generation, question disambiguation -- to gain in interpretability and effectiveness, LLM4CS aggregates several rewrites and answers within the representation space of the retrieval models~\cite{mao2023largelanguagemodelsknowllm4cs}, and MQ4CS focuses on multi-aspect query rewrites~\cite{abbasiantaeb2024generateretrieveconversationalresponse}. However, context modeling using \acp{LLM} is too computationally costly, making it impossible for production. 

\header{Distillation in \ac{CS}.} Lots of research aims at modeling the conversational context via learning to represent the conversation based on gold rewritten queries. Existing methods achieve this goal by taking the representation of the rewritten gold query as a teacher model and learning to distill the representation~\cite{10.1145/3404835.3462856,hai2024cospladecontextualizingspladeconversational,10.1145/3543507.3583265,mo2024aligningqueryrepresentationrewritten}. 
ConvDR~\cite{10.1145/3404835.3462856} learns the mapping between full conversation and human rewrites representations by minimizing an \ac{MSE} loss on the CLS tokens of both, performing well for dense bi-encoders. 
QRACDR~\cite{mo2024aligningqueryrepresentationrewritten}, also proposes a similar distillation, with several new \ac{MSE} terms, between documents and queries to improve the representations and better align with the contrastive nature of the task.
We differentiate from these works as we inspire from contrastive margin distillation (Margin-MSE~\cite{hofstatter2020improving}) and relax the query rewriting distillation by learning based on the relevance scores (i.e., the dot product of the query representation and a document), rather than the representation itself. This way, not only do we enable converging towards indefinite possible query representations, but also we take advantage of the contrastive nature of ranking by aligning the rewrite task to retrieval. Furthermore, we are not bound to one query or one teacher.

\header{Learned sparse representation.}
As \ac{LSR} gained popularity in ad-hoc retrieval~\cite{lassance2024spladev3,lassance2023naver,DBLP:conf/trec/LassanceC22}, SPLADE architectures were proposed for conversational passage retrieval~\cite{hai2024cospladecontextualizingspladeconversational,10.1145/3543507.3583265,lupart2024irlab}, to benefit their interpretability and robustness properties~\cite{spladev1,10.1145/3477495.3531857/splade++,adversarial}. In coSPLADE~\cite{hai2024cospladecontextualizingspladeconversational}, the authors use an \ac{MSE} loss between full conversation and human rewrite representations of the sparse bag-of-words representations (of dimension 30k), and show promising performance. In the meantime, LeCoRe~\cite{10.1145/3543507.3583265} also aims to distill human rewrites on sparse representations, through intermediate embedding layers and by filtering a maximum of dimensions, to avoid the \ac{MSE} on large dimensionality vectors. Distilling sparse representations is, however, challenging, as it requires controlling which dimensions should be activated, often restricting the potential of the sparse representations. Additionally, using a MSE is problematic since it treats each dimension separately, neglecting the significance of the associated dimension. These issues limit the effectiveness of sparse retrieval in previous work. In our work, by only distilling end scores, we give the model complete freedom to learn which dimensions to activate. We also investigate sparsity further, including methods to control it within CS. 

\section{Proposed Method}
In this section, we first recall notations from the \ac{CS} and \ac{LSR} fields, before presenting  DiSCo and our relaxed distillation objective.

\subsection{Preliminaries}
\header{Notation.}
We consider a set of conversations between a user and a system, each composed of multiple turns. At turn $n$, we have access to previous queries and answers, together with the final user utterance $q_n$. 
We denote the full conversational context as $q_{conv}$, separated with \texttt{[SEP]} tokens, and $q_{rw}$ as the gold rewritten query. $q_{rw}$ resolves various complexities such as ellipsis and ambiguities of the conversation and can be generated by either human or LLM.
$$
q_{conv} = q_n, a_{n-1}, q_{n-1}, ..., a_0, q_0~.
$$

Furthermore, similarly to other approaches in CS~\cite{10.1145/3404835.3462856,hai2024cospladecontextualizingspladeconversational,10.1145/3543507.3583265,mo2024aligningqueryrepresentationrewritten} we rely on two already trained encoder models $E_q$ and $E_d$, for queries and documents representations.
These backbone models are trained on a large-scale IR dataset with regular short-form queries extracted from search engine logs.
We also follow the assumption made in \ac{CS} that $E_d$ is already good at representing documents and doesn't need further fine-tuning, only $E_q$ needs further fine-tuning to adapt to long conversational contexts, noted $\tilde{E}_q$. 
This can be done with a contrastive InfoNCE \cite{infonce} loss on the CS datasets (e.g., convANCE, convSPLADE), or with a distillation loss (e.g., ConvDR and other distillation approaches such as \spladis).

\header{Learned Sparse Retrieval.} \label{sparse_retrieval_recall}
We rely on the SPLADE architecture \cite{spladev1,formal2021splade}, based on the BERT pretrained transformers~\cite{DBLP:journals/corr/abs-1810-04805/bert,lin2020IRBertReview}. SPLADE makes use of the Masked Language Modeling embedding layer of BERT to create sparse representations, where each dimension corresponds to a token of the vocabulary. $E_{q}$, $E_{d}$ and $\tilde{E}_{q}$ are the MLM outputs of the model, of dimension 30k (vocabulary dimension). To control the sparsity of the model, the authors propose to use a regularization loss~\cite{paria2020minimizing}, 
The sparsity of the models can be controlled with the two hyperparameters $\lambda_q$ and $\lambda_d$, as weights for the regularization loss. As we rely on the same SPLADE model, we also included this loss. 

\begin{figure}[t]
    \centering
    \includegraphics[width=\linewidth]{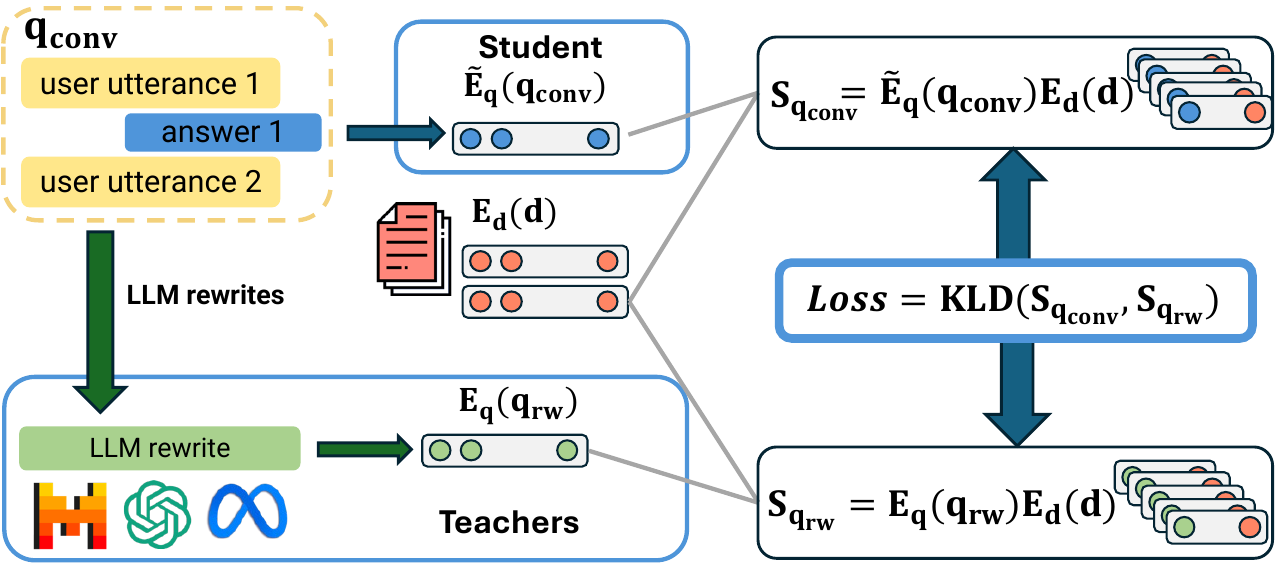}
    \caption{\spladis, as the Distillation of LLM rewritten queries through a contrastive objective. Previous works distilled representations themselves, while our approach distills similarities with documents from the corpus, relaxing the learning objective.}
    \label{fig:pres}
\end{figure}

\subsection{DiSCo}

We describe our distillation process in Figure~\ref{fig:pres}. The conversation is passed through the teacher model, rewriting the last user utterance, and computing similarities with several documents from the corpus. Meanwhile, the student encodes the full conversation and scores it with those documents. The final training objective for the student is to match the teacher similarity scores. This relaxes the previous objective, which distilled representations directly. 

\header{Representation distillation.}
Existing approaches~\cite{10.1145/3404835.3462856,hai2024cospladecontextualizingspladeconversational,10.1145/3543507.3583265,mo2024aligningqueryrepresentationrewritten} distill query rewrites on the representation space, with a direct constraint on the representation. The goal is to have the representations of the full conversations converging toward the representations of the rewritten queries:
\begin{equation}
\tilde{E_{q}}(q_{conv}) \rightarrow E_q(q_{rw})~.
\label{eq:before}
\end{equation}

\begin{figure}[t]
    \centering
    \includegraphics[width=0.9\linewidth]{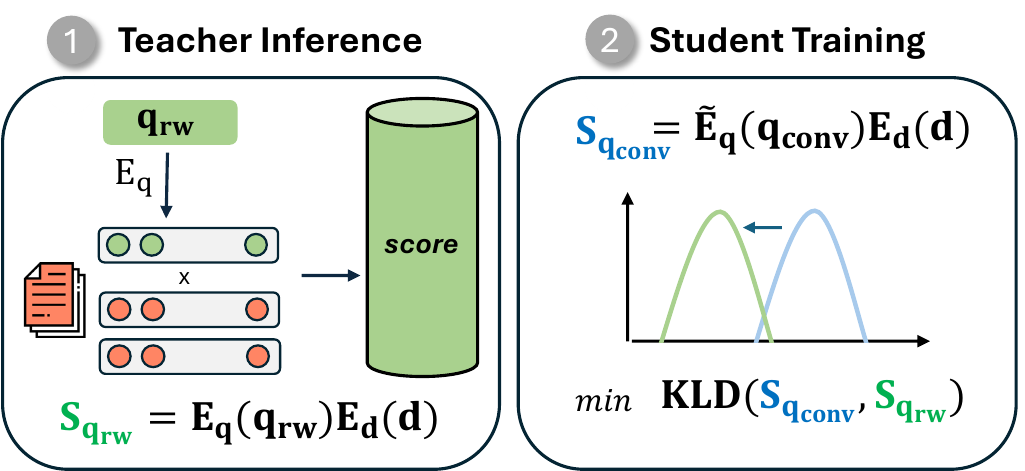}
    \caption{Distillation process. The first step stores scores from the rewritten queries with documents from the corpus. Then the student query encoder $\tilde{E}_q$ is trained to reproduce the output scores of the teacher.}
    \label{fig:distil_explained}
\end{figure}

\header{Relaxation of the distillation.}
In our work, we propose to distill the scores rather than the representations.
This allows for a relaxation of the training objective. 
Instead of forcing $\tilde{E}_{q}(q_{conv})$ to converge to $E_{q}(q_{rw})$, we allow the entire hyperplane on which the similarity for $q_{conv}$ and $q_{rw}$ with an anchor document $d$ is equal:
\begin{equation}
\tilde{E}_{q}(q_{conv}) \rightarrow \{ X \in \mathbb{R}^h \mid X^\top E_d(d) = s_{q_{rw}}\}~,
\label{eq:after}
\end{equation}
with targeted scores $s_{q_{rw}}=E_q(q_{rw})^\top E_d(d)$, and $h$ the dimension of the representation space.
More intuitively, while the existing distillation loss functions bound the model to converge to one single embedding~\cite{10.1145/3404835.3462856}, our loss allows for infinite possible optimum embeddings, as long as they have the same dot product with the target relevant document embedding (i.e., any point in the hyperplane). 
Such distillation on scores can be trained with a Kullback Leibler Divergence loss~\cite{Kullback1951OnIA} on the distribution of scores:
\begin{equation}
\mathcal{L}_{\mathbf{KLD}} = D_{\text{KL}}(\mathcal{S}_{q_{rw}} \,||\, \mathcal{S}_{q_{conv}})~,
\label{eq:kld}
\end{equation}
where $\mathcal{S}_{q_{rw}}$ and $\mathcal{S}_{q_{conv}}$ are the distributions of similarity scores within the batch. It differs from the existing learning objective which is achieved with an \ac{MSE} loss on each dimension of the vector representations independently~\cite{10.1145/3404835.3462856,hai2024cospladecontextualizingspladeconversational}:
\begin{equation}
\mathcal{L}_{\mathbf{MSE}} = \mathbf{MSE} ( \: \tilde{E}_q(q_{conv}),\: E_q(q_{rw}) )~.
\label{eq:mse_rep}
\end{equation}
This new $\mathcal{L}_{\mathbf{KLD}}$ loss includes the benefit of the contrastive objective, anchoring itself on relevant and hard documents from the corpus. Also note that in the previous optimization, the final loss was the sum of the distillation MSE loss with the contrastive InfoNCE loss, while we only use a single contrastive distillation KLD loss, unifying both objectives into one distillation loss.
Figure~\ref{fig:distil_explained} illustrates the distillation process. First, the teacher retrieves documents using the rewritten queries and stores the similarity scores. Then, the student model encodes the full conversation, learning to have equal similarities to the teacher on these documents, by minimizing the KLD distillation loss. This defines the final learning objective of our \spladis models, as minimizing the distributions of scores between the student and teacher models.

\header{Hard negatives.} As our new training objective from Equation~\ref{eq:kld} is now contrastive, we can benefit from hard negative mining. This was not possible with the previous distillation objective, as it only uses conversation representations, independently from documents.

Thus applied to the case of multiple negatives per query, the target space from Equation \ref{eq:after} becomes a subspace of $\textbf{(h-N)}$ dimensions. This still keeps a higher degree of freedom, since the number of negatives is much lower than the dimension of the embedding space ($h>>N$).
\begin{equation}
\{ X \in \mathbb{R}^h \mid X^\top E_d(d_i) = s_{q_{rw}}^i \ \forall i \in [1, N] \}~,
\end{equation}
with $s_{q_{rw}}^i = E_q(q_{rw})^\top E_d(d_i)$ being the scores from multiple hard negatives. These negatives are mined during the Teacher Inference step from Figure~\ref{fig:distil_explained}, further improving training objective~\cite{DBLP:journals/corr/abs-2007-00808/ance,10.1145/3477495.3531857/splade++}.
 
\header{Teacher Models.} 
An important component of distillation relies on the choice of the teacher. The teacher is first used to rewrite the conversation utterance, into $q_{rw}$, which will be encoded $E_q(q_{rw})$ and distilled through the similarities scores $s_{q_{rw}}$. We use multiple LLMs, together with human rewrite (when available), as teachers. Having several teachers is motivated by the well-established effectiveness of ensembling strategies, where stronger teachers would provide higher-quality training signals, leading to better student model performance. This results in several trained \spladis students, depending on the teacher used. Below we list all the teachers:
\begin{itemize}[nosep]
    \item \textbf{T0}: T5QR 
    \item \textbf{T1}: LlamaQR 
    \item \textbf{T2}: MistralQR 
    \item \textbf{T3}: HumanQR
\end{itemize}

Furthermore, another property of the proposed distillation is that we can distill from multiple teachers, making the distill from several \acp{LLM} or humans possible. We do so by averaging the similarity scores of the $T$ teachers, each having a different rewrite, thus a different similarity with the documents. We experimented with mean, min, and max aggregation methods, observing no significant differences in performance, and thus decided to use the more simple mean aggregation. Overall, the distilled score is:
\begin{equation}
s_{q_{rw_{all}}} = \frac{s_{q_{\text{rw}_1}} + s_{q_{\text{rw}_2}} + \cdots + s_{q_{\text{rw}_T}}}{T}~.
\end{equation}

The final learning objective of our \spladis multi-teacher is the same as for the regular \spladis (i.e., KLD from Equation~\ref{eq:kld}), but distilling the average score of the set of teachers, instead of using a single similarity from a unique teacher. 

\header{Student-Teacher Fusion.} 
Finally, we propose a fusion of the teacher and the student model, combining the ranked lists of both at inference. This is achieved using an average normalized score
over the ranked list of both models.

\section{Experiments Design}

\subsection{Datasets and Metrics}

\header{Datasets.}
We evaluate our method's effectiveness on various conversational passage retrieval datasets, detailed in Table~\ref{tab:dataset-stat}, namely:
\begin{itemize}[nosep,leftmargin=*]
    \item \textbf{QReCC} \cite{anantha-etal-2021-open} is built upon ORConvQA~\cite{10.1145/3397271.3401110/orconvqa}, NQ~\cite{47761/nq} and TREC CAsT 2019 \cite{10.1145/3397271.3401206/cast19}. It features 13K conversations and 80K turns created by human annotators around existing documents.

    \item \textbf{TopiOCQA}~\cite{adlakha-etal-2022-topiocqa} distinguishes itself from QReCC focusing on topic switches, where every Wikipedia hyperlink is considered a topic switch in a conversation built around that Wikipedia topic. It features ~4K conversations and ~48K turns.  

    \item \textbf{TREC CAsT 2020, 2022 }~\cite{Dalton2020CAsT2T,Owoicho2022TRECC2} and \textbf{TREC iKAT 2023}~\cite{10.1145/3626772.3657860} are smaller conversational datasets, but carefully hand-crafted to include various conversational complexities. They are coupled with high-quality relevance judgements done by NIST assessors.
\end{itemize}

\begin{table}[t]
    \centering
    \caption{Statistics of the datasets.}
    \begin{tabular}{llllc}
    \toprule  
    \textbf{Dataset} & \textbf{Split} & \textbf{\# Conv.} & \textbf{\# Turns} & \textbf{Collection} \\
    \midrule
    \multirow{2}{*}{QReCC} & Train & 10,823 &  63,501 & \multirow{2}{*}{\phantom{0} 54M} \\
    & Test & \phantom{0}2,775 & 16,451 &  \\
    \midrule
    \multirow{2}{*}{TopiOCQA} & Train & \phantom{0}3,509 & 45,450 & \multirow{2}{*}{\phantom{0}25M} \\
    & Test & \phantom{00,}205 & \phantom{0}2,514 &  \\
    \midrule
    TREC CAsT 20 & Test & \phantom{00,0}25 & \phantom{00,}216 & \phantom{0}38M \\
    TREC CAsT 22 & Test & \phantom{00,0}18 & \phantom{00,}205 & 138M \\
    TREC iKAT 23 & Test & \phantom{00,0}25 & \phantom{00,}176 & 116M \\
    \bottomrule
    \end{tabular}
    \label{tab:dataset-stat}
\end{table}

\header{Effectiveness Metrics.}
We report the results in terms of the main IR metrics~\cite{10.1145/3404835.3462856,10.1145/3543507.3583265}, as well as official TREC CAsT and iKAT ones~\cite{10.1145/3397271.3401206/cast19,10.1145/3626772.3657860}. Metrics are Mean Reciprocal Rank (MRR), Normalized Discounted Cumulative Gain (nDCG)~\cite{ndcg} at 3. As we focus on first-stage retrieval, we also include Recall at different ranks~\cite{recall_neural_ir}. We determine the statistically significant differences by doing a two-sided paired t-test with Bonferroni correction at 95\% confidence ($p < 0.05$).

\header{Efficiency Metrics.}
In terms of inference efficiency, we report the number of \textit{LLM inference calls} when used for query rewriting, as it is a strong indicator of inference latency. As an example, it takes an average of 4.4 seconds to generate 64 tokens on Llama 3.1, on A100 GPU, while the typical dual-encoder retrieval latency is in the range of 100 milliseconds on CPU~\cite{10.1145/3477495.3531833/efficient,10.1145/3539618.3591941/pruning,twostepsplade}.

We also provide Rewrite and Retrieval efficiency, of several models. This is performed per query, averaged over the entire test set. Rewrite is the latency for rewriting, while retrieval is both query encoding and search through the inverted index (in the case of SPLADE models). We used a 4th AMD EPYC CPU and an A100 GPU.

We also report the FLOPs~\cite{paria2020minimizing,formal2021splade} values, a metric for efficiency in learned sparse retrieval, which intuitively estimates the number of floating point operations between a query and a document in an inverted index. FLOPs can be computed as follows:
\begin{equation}
    FLOPs = \mathbb{E}_{q,d}[\sum_{j \in V} p_j^{(q)} p_j^{(d)}]~,
    \label{eq:flops}
\end{equation}
where $p_j^{(q)}$ and $p_j^{(d)}$ are the probabilities of activation of the $j^{th}$ token in the vocabulary, resp. in query and document representations, over which we average on the dataset distribution.

\subsection{Baselines}

We compare our \spladis model with a wide range of competitive methods, consisting of query rewriting, supervised fine-tuned, and distillation-based methods, which we list below.

\header{Query rewriting methods.}
\emph{SPLADE-[T5/Llama/Mistral/Human]QR} does retrieval using the SPLADE ad-hoc retrieval model trained on MS MARCO~\cite{DBLP:journals/corr/NguyenRSGTMD16}. As input, we pass the rewritten query using T5~\cite{DBLP:journals/corr/abs-1910-10683/T5}, Llama 3.1~\cite{dubey2024llama3herdmodels}, Mistral~\cite{jiang2023mistral7b}, or gold human rewrites\footnote{We used the latest Llama \texttt{meta-llama/Meta-Llama-3.1-8B-Instruct}, Mistral \texttt{mistralai/Mistral-7B-Instruct-v0.2} and T5  \texttt{castorini/t5-base-canard}}. \emph{SPLADE no rewrite} is the same ad-hoc retrieval model without rewrite, on the original conversation. 
\emph{IterCQR}~\cite{jang-etal-2024-itercqr}, \emph{CHIQ-Fusion}~\cite{mo2024chiqcontextualhistoryenhancement}, and \emph{LLM4CS}~\cite{mao2023largelanguagemodelsknowllm4cs} are state-of-the-art query rewriting baselines; however, they require multiple LLM calls at inference, which puts them under a high disadvantage. For \emph{LLM4CS}, we reproduced their best setting (RAR, Mean aggregation from $N=5$, CoT, GPT-4).

\header{Supervised fine-tuned methods.}
\emph{convSPLADE}~\cite{10.1145/3543507.3583265} and \emph{convANCE}~\cite{10.1145/3543507.3583265} are two methods fine-tuned using the InfoNCE loss on the conversational contrastive labels. Their input is the whole conversational context.
We reproduce convSPLADE for out-of-domain retrieval using the same hyperparameters as in the original paper.

\header{Distillation-based method.}
\emph{LeCoRe}~\cite{10.1145/3543507.3583265}, \emph{QRACDR}~\cite{mo2024aligningqueryrepresentationrewritten} and \emph{ConvDR}~\cite{10.1145/3404835.3462856}\footnote{We did not manage to reproduce the results reported in the original papers, as our reproduced results were much lower. Therefore, we report the numbers in the original papers. Therefore, we are not able to perform any significance tests with these models.}
     all learn to distill the gold human rewrite representations and usually combine the distillation loss with an InfoNCE. Like our \spladis, these models do not rely on LLM calls either, making them comparable. \emph{ConvDR} and \emph{QRACDR} are both dense approaches, using latent representations, while \emph{LeCoRe} uses learned sparse representations. All of them distill query representations, while \spladis the similarity scores, as a relaxation of their distillation method.

\begin{table}[t]
  \caption{One-shot prompt used for rewriting with the LLMs teachers. The example is taken out of the QReCC dataset.}
  \label{tab:prompts}
    \begin{lstlisting}
    # Instruction: I will give you a conversation between a user and a system. You should rewrite the last question of the user into a self-contained query.
    # Example 1:
    # Context:
    user: Tell me about the benefit of Yoga?
    system: Increased flexibility, muscle strength.
    # Please rewrite the following user question:
    Does it help in reducing stress?
    # Re-written query:
    Does Yoga help in reducing stress?
    # Example 2:
    # Context:
    <ctx>
    # Please rewrite the following user question:
    <utterance>
    # Re-written query:
    \end{lstlisting}
\end{table}

\begin{table*}[t!] 
  \caption{In-Domain Performance on TopiOCQA and QReCC. \spladis multi-teach for TopiOCQA is the combination of $T_1$ and $T_2$, and QReCC uses $T_2$ and $T_3$. Hyperscripts $\dagger$ are paired t-test $p<0.05$ comparing multi-teachers with single-teacher \spladis. RW denotes the LLM/human rewriting method used as input to the model. FC refers to the models that do not use any rewriting at inference and just take the Full Context as input. \includegraphics[width=0.3cm]{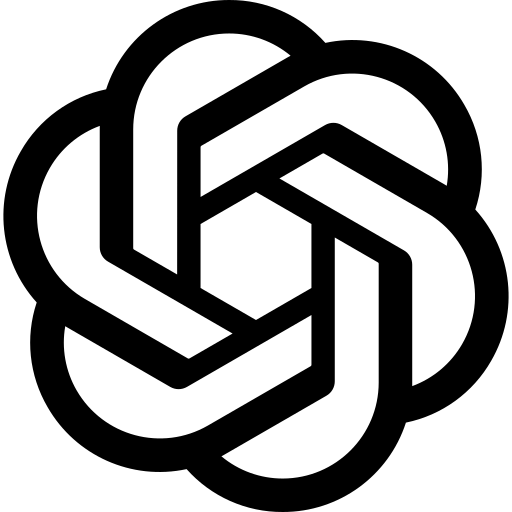} denotes the number of LLM calls used at inference.}
  \label{main_table}
  \begin{tabular}{lllP{1cm}cccccccccc}
    \toprule
    &\multirow{2}{*}{\textbf{Method}} & \multirow{2}{*}{\textbf{RW}} & \multirow{2}{*}{\includegraphics[width=0.5cm]{figures/gpt-icon.png}} &\multicolumn{4}{c}{\textbf{TopiOCQA}} &&  \multicolumn{4}{c}{\textbf{QReCC}} \\    
    \cmidrule{5-8} \cmidrule{10-13}
     &&&& R@100 & R@10 & MRR & nDCG@3 &&  R@100 & R@10 & MRR & nDCG@3 \\
     \midrule
    
     \multirow{9}{*}{\rotatebox[origin=c]{90}{\textbf{Query Rewriting}}} & SPLADE no rewrite & FC & 0  &   0.472 & 0.258 & 0.155 & 0.141   && 0.840 & 0.673 & 0.485 & 0.459 
 \\
      & SPLADE HumanQR ($T_3$)& Human & 0 &   -& -& - & -&& 0.912 & 0.714 & 0.448 & 0.433  \\
     \cmidrule{2-13}
     &SPLADE T5QR ($T_0$) &T5&1 & 0.667 & 0.501 & 0.321 & 0.314 &&  0.840 & 0.617 & 0.382 & 0.366 \\
     &SPLADE LlamaQR ($T_1$)&Llama 3 & 1 & 0.761 &  0.572 &  0.365 & 0.352 &&  0.826 & 0.613 & 0.377 & 0.360 \\
     &SPLADE MistralQR ($T_2$) & Mistral 2 & 1 & 0.759 & 0.591 & 0.366 & 0.356&&  0.884 & 0.668 & 0.424 & 0.409  \\
    \cmidrule{2-13}
     &IterCQR \cite{jang-etal-2024-itercqr} & GPT-3.5 & 1  &  0.620 & 0.426 &  0.263  &  0.251 && 0.841 & 0.655 & 0.429 & 0.402   \\
     &LLM4CS \cite{mao2023largelanguagemodelsknowllm4cs} & GPT-3.5 & 5 &- & 0.433 & 0.277 & 0.267  &&  - & 0.664 & 0.448 & 0.421 \\ 
     &CHIQ FT \cite{mo2024chiqcontextualhistoryenhancement}&T5 & 1 & - & 0.510 &  0.300 &  0.289 && - &  0.576 & 0.369 & 0.340  \\
     &CHIQ-Fusion \cite{mo2024chiqcontextualhistoryenhancement} & Llama 2 & 6 & - & 0.616 & 0.380 & \underline{0.370} && - & 0.707 & 0.472 & 0.442  \\
    \midrule

    \multirow{2}{*}{\rotatebox[origin=c]{90}{\textbf{SFT}}} & convANCE  \cite{10.1145/3543507.3583265}& FC & 0 & 0.710 & 0.430 & 0.229 & 0.205 &&  0.872 & 0.715 & 0.471 & 0.456  \\ 
     & convSPLADE  \cite{10.1145/3543507.3583265} & FC & 0 & 0.720 & 0.521 & 0.295 & 0.307   && 0.878   &  0.699 & 0.500 & 0.466  \\
    \midrule
    
    \multirow{7}{*}{\rotatebox[origin=c]{90}{\textbf{Distillation}}} & ConvDR~\cite{10.1145/3404835.3462856} & FC &0 &  0.611 & 0.435 & 0.272 & 0.264 && 0.778 & 0.582 & 0.385 & 0.357 \\
     & QRACDR~\cite{mo2024aligningqueryrepresentationrewritten} & FC & 0 &  0.758  &  0.571 & 0.377 & 0.365 && 0.897  &  \underline{0.748} & \textbf{0.516} & \textbf{0.516}  \\
     & LeCoRe~\cite{10.1145/3543507.3583265} & FC & 0 & 0.735 &0.543 & 0.320 & 0.314  && 0.897 &  0.739 &\underline{0.511} & 0.485    \\
    \cmidrule{2-13}
     &\spladis T5 & FC & 0 & 0.834&	0.617	&0.363&0.345&&0.917&0.719&0.456&0.442\\
     &\spladis Mistral & FC & 0 & \underline{0.842} & \underline{0.634} &  \underline{0.387} & \underline{0.370} &&  0.925 & 0.743 & 0.489 & 0.477  \\
     &\spladis Human & FC & 0 &  - &  - &  - &  -  &&  \underline{0.927} & 0.741 & 0.483 & 0.470  \\
     &\spladis multi-teach & FC & 0 & \textbf{0.859}$^{\dagger}$&  \textbf{0.640} & \textbf{0.390}&  \textbf{0.375} &&  \textbf{0.928} & \textbf{0.754}$^{\dagger}$ & 0.498$^{\dagger}$ &  \underline{0.487}$^{\dagger}$  \\
    \bottomrule
    
  \end{tabular}
\end{table*} 

\subsection{Implementations details}
We use the SPLADE++~\cite{10.1145/3477495.3531857/splade++}\footnote{\texttt{naver/splade-cocondenser-ensembledistil}~\cite{10.1145/3477495.3531857/splade++}} checkpoints from Huggingface~\cite{wolf2020huggingfaces} for all our SPLADE models. 
To further finetune on the \ac{CS} task, we fine-tune for 5 epochs, with a learning rate of $2e^{-5}$, a max sequence length for queries of 64 tokens, and 100 tokens for answers, with a total limit of 256 tokens for the full conversational context. We also use a 256-token limit for passages. We use a batch size of 10, with 16 negatives per query, and in-batch negatives~\cite{lin-etal-2021-batch}. Default experiments with SPLADE use FLOPS and L1 regularization, as in the original code, with values $\lambda_q=1e^{-3}$, $\lambda_d=5e^{-4}$. 
We use mixed precision and fp16 to maximize memory use.
For sparse retrieval, we use inverted indexes based on the \texttt{numba} library and \texttt{pyserini} \cite{Lin_etal_SIGIR2021_Pyserini} together with \texttt{Pytorch}~\cite{paszke2019pytorch}. The fusion method uses the ranx library~\cite{ranx}.
All teachers use In-Context learning to generate the rewrites.
We provide the one-shot prompt used for rewriting with the LLM Teacher models in Table~\ref{tab:prompts}.  We experimented in both zero-shot and one-shot, but decided to use one-shot to improve the quality of the teacher models. 
Given the high efficiency of our fine-tuning, we were able to run the experiments on a single A100 GPU with 40GB of GPU memory.

\section{Results}

In this section, we present the results of our \spladis and the proposed distillation for in-domain and out-of-domain retrieval, together with an analysis of the multi-teacher distillation and the efficiency-effectiveness trade-off. 

\header{Research questions.} We aim to answer the following questions:
\begin{enumerate}[label=\textbf{RQ\arabic*}]
    \item Would relaxing the distillation training objective improve conversational sparse retrieval in-domain performances? \label{rq1}
    \item How would the relaxation affect the out-of-domain generalization capacities of the models? \label{rq2}
    \item How does the distillation of multiple \ac{LLM} teachers compare to single teacher distillation? \label{rq3}
    \item How does the relaxed distillation affect the efficiency and sparsity of the student models? \label{rq4}
\end{enumerate}

\begin{table*}[t]
  \caption{Zero-shot Performance on Out-Of-Domain. \spladis multi-teach uses both $T_2$ and $T_3$ (Mistral and Human teachers). \spladis Fusion is the fusion of the SPLADE MistralQR with \spladis multi-teach. FC refers to the models that do not use any rewriting at inference and just take the Full Context as input. \includegraphics[width=0.3cm]{figures/gpt-icon.png} denotes the number of LLM calls used at inference.}
  \label{zs_table}
  \resizebox{\textwidth}{!}{
  \begin{tabular}{lllP{0.4cm}cccccccccccc}
    \toprule
    &\multirow{2}{*}{\textbf{Method}} & \multirow{2}{*}{\textbf{RW}} & \multirow{2}{*}{\includegraphics[width=0.45cm]{figures/gpt-icon.png}} &\multicolumn{3}{c}{\textbf{CAsT 2020}}&&\multicolumn{3}{c}{\textbf{CAsT 2022}}&&\multicolumn{3}{c}{\textbf{iKAT 2023}} \\
    \cmidrule{5-7} \cmidrule{9-11} \cmidrule{13-15}
     && & & R@100 & MRR & nDCG@3& &R@100 & MRR & nDCG@3& &R@100 & MRR & nDCG@3\\
    \midrule
    \multirow{9}{*}{\rotatebox[origin=c]{90}{\textbf{Query Rewriting}}}& SPLADE HumanQR&Human&0  &0.646&0.636& 0.475 & & 0.422  & 0.590 & 0.423 && 0.285  & 0.359 & 0.262 \\
    \cmidrule{2-15}
     &SPLADE T5QR&T5&1 &    0.479 &  0.477& 0.332 &&         0.226 &0.355& 0.218 & & 0.115 &  0.200 &0.132\\
     &SPLADE LlamaQR&Llama 3&1 & 0.550& 0.515& 0.376 &&     0.312& 0.453& \underline{0.300} &&\underline{0.198} &0.281& 0.177\\
    & SPLADE MistralQR &Mistral 2&1 &\underline{0.572} &   0.553 & 0.403&&\underline{0.337}&\underline{0.487}&0.298 & &0.178  & \underline{0.291} & \textbf{0.194} \\
    \cmidrule{2-15}
    & LLM4CS \cite{mao2023largelanguagemodelsknowllm4cs}&GPT 3.5&5 &0.489&0.615& \textbf{0.455}&&-&-&-&&-&-&-\\  
    & LLM4CS (ours)&GPT 4&5 &0.504&\textbf{0.618}& 0.444&&0.283&0.425&0.272&&0.133&0.154&0.099\\  
    & CHIQ FT \cite{mo2024chiqcontextualhistoryenhancement}&T5 &1 & -& 0.463& 0.316 &&-&-&-&&-&-&-\\
    & CHIQ Fusion \cite{mo2024chiqcontextualhistoryenhancement}&Llama 2&6 & -& 0.540&0.380&&-&-&-&&-&-&- \\
    &\spladis Fusion &Mistral 2&1 &  \textbf{0.611}  & \underline{0.566} & \underline{0.425} && \textbf{0.379} & \textbf{0.578} & \textbf{0.384} && \textbf{0.201}  & \textbf{0.297} & \underline{0.192} \\
    \midrule
    \midrule
    \multirow{2}{*}{\rotatebox[origin=c]{90}{\textbf{SFT}}}& \multirow{2}{*}{convSPLADE
    (ours)}&\multirow{2}{*}{FC}&\multirow{2}{*}{0} & \multirow{2}{*}{0.446}&\multirow{2}{*}{0.338}&\multirow{2}{*}{0.234}&&\multirow{2}{*}{0.274} &\multirow{2}{*}{0.382}&\multirow{2}{*}{0.227}&& \multirow{2}{*}{0.101}& \multirow{2}{*}{0.144} &\multirow{2}{*}{0.085 }\\
    \\
    \midrule
    \multirow{5}{*}{\rotatebox[origin=c]{90}{\textbf{Distillation}}}& QRACDR \cite{mo2024aligningqueryrepresentationrewritten}&FC &0&  0.324  & 0.442&0.303&&-&-&-&&-&-&- \\
    & LeCoRe \cite{10.1145/3543507.3583265} &FC&0&  0.467  &  -&0.290&&-&-&-&&-&-&-\\
    \cmidrule{2-15}
    & \spladis Mistral &FC&0  & 0.519  & \underline{0.457} & \underline{0.341}& &\underline{0.322}&0.463&0.287&&0.135&\underline{0.193}&\underline{0.126}\\
     & \spladis Human &FC&0 & \underline{0.523}  & 0.455 & 0.339& &0.314 & \underline{0.490} & \underline{0.308} && \textbf{0.151}  & \textbf{0.202} & \textbf{0.131} \\
    & \spladis multi-teach &FC&0 & \textbf{0.531} & \textbf{0.483} & \textbf{0.353} & & \textbf{0.334} & \textbf{0.512} & \textbf{0.322} &    & \underline{0.147}  & 0.192 &0.125 \\
    \bottomrule    
  \end{tabular}
  }
\end{table*}

\subsection{Performance Comparison}
In this first subsection, we aim to answer \ref{rq1} and \ref{rq2} by comparing the performance of our proposed \spladis with state-of-the-art query rewriting, and distillation-based methods. We first compare the performance of \spladis in the in-domain and out-of-domain retrieval setups in Tables~\ref{main_table} and \ref{zs_table}, respectively. The tables report the performance of diverse baselines in terms of various metrics, as well as the number of LLM calls every model requires at inference. This is an important factor while comparing the performance of conversational retrieval models for various reasons, namely, 
\begin{enumerate*}[label=(\roman*)]
    \item LLM call leads to considerable inference latency, delaying inferencing from hundreds of milliseconds to seconds; and
    \item LLM-based methods take advantage of the vast parameter size and knowledge learned by the training of the LLMs, making their comparison to other smaller models unfair.
\end{enumerate*}

\header{In-domain retrieval.}
Trying to address \ref{rq1}, we report in Table~\ref{main_table} in-domain performance on \topiocqa and \qrecc. Looking at the results, we do not observe a big gap between the rewriting and distillation-based methods, even though the rewriting-based methods make use of the LLM knowledge in the rewriting phase. Distillation-based methods (i.e., ConvDR, QRACDR, LeCoRe) even outperform the rewriting-based methods while being more efficient too, as they have zero LLM calls. 

Besides, our \spladis outperforms all rewriting- and distillation-based baselines with a large margin, showing that relaxing the distillation constraint to learn the similarity score, rather than the representations leads to further improvements. In particular, we see that \spladis Mistral manages to outperform the learned sparse baseline, LeCoRe, by 10 and 2.8 points on \topiocqa and \qrecc resp.\ in terms of Recall@100.

\header{Out-of-domain retrieval.}
To further study the effectiveness of our proposed distillation approach, and study its generalizability, we report in Table~\ref{zs_table} the results in the out-of-domain retrieval setting on TREC CAsT 20,22 and iKAT 23. Addressing \ref{rq2}, we see that, unlike the in-domain setting, there is a considerable gap between the rewriting-based and distillation-based methods when it comes to out-of-domain retrieval. On average, rewriting-based methods perform 19\% better than distillation-based methods because they mainly rely on the massive parameter size and knowledge of LLMs, leading to high generalizability. 
However, as mentioned earlier making LLM calls on inference puts the models at a high disadvantage because of the high latency. 

Focusing on distillation-based approaches, our \spladis outperforms other methods by a large margin, showing the remarkable generalizability of our proposed relaxed distillation. In particular, we see that \spladis Human outperforms QRACDR by 3.8 points in terms of nDCG@3, and LeCoRe by 5.5 points in terms of Recall@100, both on CAsT 2020. Note that those two models are two state-of-the-art models in the same zero-shot settings, trained on QReCC and evaluated on CAsT 2020. It is also noteworthy that even though our distillation-based approach is not able to outperform some of the rewriting-based (LLM-based) methods, our \spladis-Fusion, which is a fusion method based on LLM rewriting and distillation manages to outperform all rewriting-based methods by a large margin in terms of Recall@100, reaching SotA conversational passage retrieval performance. When comparing closely with the SotA LLM4CS model, we note that while LLM4CS takes advantage of 5 GPT-4 calls, our \spladis multi-teach model outperforms it in terms of recall on all out-of-domain datasets. Comparing the precision-oriented metrics, we see that \spladis multi-teach outperforms LLM4CS on all datasets, except TREC CAsT 2020. This dataset was the second edition of TREC CAsT and is considered to be simpler than both 2022 and 2023 versions, showing the performance of our model on complex information needs. Also, considering that we focus on the retrieval task, we consider recall to be preferable, as reranking can be added post-retrieval on a smaller set of documents~\cite{recall_neural_ir,dejean2024rerank}. 

\header{Distillation-rewriting fusion.}
Furthermore, note in Table~\ref{zs_table} that \spladis-Fusion is the fusion of our \spladis and SPLADE MistralQR, as the fusion of the student with the teacher. Although SPLADE MistralQR exhibits high performance (e.g., 0.57 and 0.33 in terms of Recall@100 resp. on CAsT 2020, 2022), fusing it with the distilled \spladis leads to further significant improvements (4 points increase on both datasets). This indicates that relaxing the distillation process to the similarity score enables the model to go beyond the knowledge of the teacher in the representation space. This gain is even stronger on the precision of CAsT 2022, where \spladis-Fusion outperforms SPLADE MistralQR by 8 points in terms of nDCG@3.

\begin{figure}[t]
    \centering
    \includegraphics[width=0.9\linewidth]{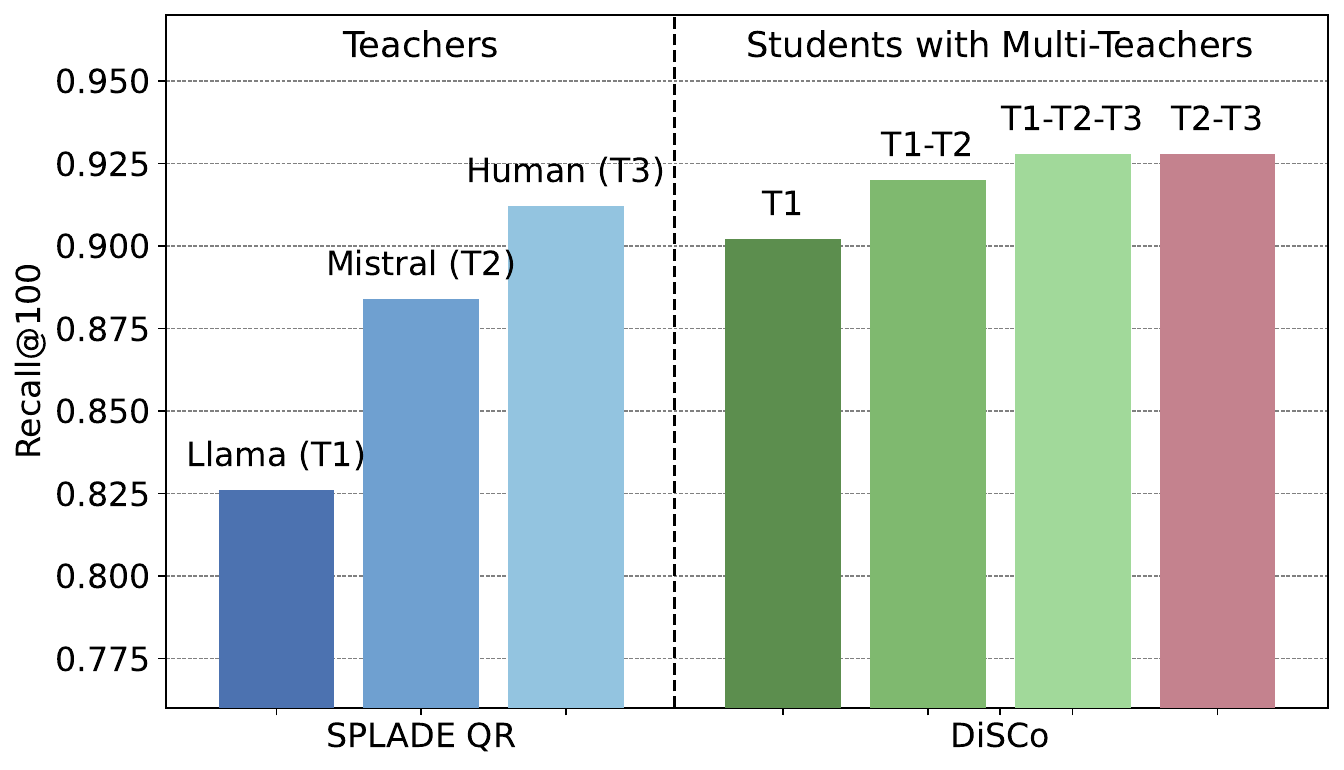}
    \caption{Teacher Selection on QReCC. (Left) SPLADE Teacher Models with different LLM QR. (Right) \spladis Students when trained with multi-teachers. Best set in red ($T_2$ and $T_3$).}
    \label{fig:barteacher}
\end{figure}

\header{Comparison with the teacher models.} Now looking at the performance of the students and teachers models in Table~\ref{main_table}, we see that the student models outperform the teachers significantly for in-domain retrieval. In particular, \spladis Human outperforms SPLADE HumanQR by 3.5 MRR points on \qrecc; and similarly with \spladis Mistral. This demonstrates that our distillation objective allows the student model to learn representations that surpass the original teacher's representations. This is possible thanks to the relaxation of the distillation, as the student can adapt and learn optimal representations based on gradient descent optimization. This was not possible with the previous distillation objective on the representation, as student models were trained to copy the representations of the teacher model.  Out-of-domain however, we see in Table~\ref{zs_table} that the student models have more difficulties outperforming the performance of the teacher models. This result on out-of-domain retrieval is, however, not fair considering that teacher model generalization relies on billions of parameters, while students on a few hundred. 
Recall also that \spladis-Fusion as the fusion of the teacher and the student was further improving performance out-of-domain.

\header{Reliance on weak teacher model.} Finally, we explore the reliance on the teacher models, by training \spladis with a weaker teacher model, such as T5. This is interesting in scenarios where we do not have access to strong teachers and also shows the robustness of the method when trained on more noisy labels. From Table~\ref{main_table}, we see that \spladis T5  performs almost on par with \spladis Mistral, while SPLADE MistralQR outperforms SPLADE T5QR by an important margin, on both \topiocqa and \qrecc. Our method is thus robust even when trained on lower-quality rewrites from the teacher.

\subsection{Multiple Teachers Distillation}

While the previous section focused on the distillation of single teachers, here we answer \ref{rq3} on the use of multiple teachers. Thus, we compare our \spladis model trained with multiple teachers.

\header{In-domain and out-of-domain retrieval.}
Trying to address \ref{rq3}, both Tables~\ref{main_table} and \ref{zs_table} include our \spladis multi-teach model trained with multiple teachers. Considering the performances of \spladis multi-teach for in-domain retrieval, we observe significant gains compared to \spladis Mistral on \qrecc. In particular, we observe a 1-point increase in terms of R@10, MRR, and nDCG@3 on \qrecc. This result remains consistent for out-of-domain retrieval, where within Table \ref{zs_table} \spladis multi-teach outperforms its single version model, by at least 2 points on all metrics, on CAsT 2020 and 2022. On iKAT 2023, however, we observe mixed results. 
Two interpretations are possible for these general gains: first, the multiple teachers and their subsequent rewrites produce word synonyms, as a form of query expansion, improving the performance, and second, as one LLM may fail to resolve the context modeling task, another can balance it, offering better robustness.

\begin{table}[t]
    \centering
    \caption{Inference Efficiency on TREC CAsT 2020. Rewrite and query encoding are measured on GPU, while the inverted index search on CPU with the Numba library. \spladis Fusion is the fusion of SPLADE MistralQR with \spladis Mistral.}
    \begin{tabular}{lccc|c}
    \toprule  
    \textbf{Efficiency (ms)} & \textbf{Rewrite} & \textbf{Retrieval} & \textbf{Total} & \textbf{R@100} \\
    \midrule
    SPLADE T5QR & 190 & 358 & 548 & 0.479 \\
    SPLADE LlamaQR & 4245 & 367 & 4612 & 0.550 \\
    SPLADE MistralQR & 2094 & 356 & 2450 & 0.572 \\
    \midrule
    convSPLADE & 0 & 356 & 356 & 0.446 \\
    \spladis Mistral & 0 & 346 & 346 & 0.519 \\
    \spladis multi-teach & 0 & 357 & \underline{357} & \underline{0.531} \\
    \midrule
    \spladis Fusion & 2094 & 356 & \textbf{2450} & \textbf{0.611} \\
    \bottomrule
    \end{tabular}
    \label{tab:latencyms}
\end{table}

\header{Progressive Improvement.}
Further addressing \ref{rq3}, Figure \ref{fig:barteacher} gives insights on the multiple teachers distillation. While the left part of the Figure gives the performances of the teachers on QReCC, the right part shows the gains when incrementally adding teachers to the distillation. 
We can also notice the important gains from the distillation of LlamaQR. In particular, \spladis Llama improves by 8 points compared to SPLADE LlamaQR in terms of Recall@100, while in comparison \spladis Mistral by only 4 points compared to his teacher. This shows that the relaxation is robust and can learn from even lower-quality rewrites.

\subsection{Effectiveness-Efficiency Trade-off}

In this subsection, we aim to answer \ref{rq4} by analyzing the effecti-veness-efficiency trade-off of the proposed \spladis models. First, we measure the efficiency of our method compared to rewriting-based methods. Then, we show the possibility of controlling the sparsity of the model through the FLOP regularization. Finally, we dive into sparsity at different depths of conversations.

\header{Efficiency.} As to explicit the inference efficiency gain of our method compared to rewrite-based approaches, we plot in Table~\ref{tab:latencyms} the efficiency in milliseconds of the models on TREC CAsT 2020. We observe from the table that while SPLADE MistralQR achieves very high performance here out-of-domain, the efficiency is very low compared to \spladis Mistral. This is because of the rewriting step, which involves 1 LLM call for MistralQR. Also note that the rewrite needs to be executed on GPU, while most of the retrieval time is a search through the inverted index on CPU. \spladis Mistral and multi-teach thus appear to have a better trade-off when it comes to efficiency. Finally, \spladis Fusion has the best performance overall, without an important computational overload compared to SPLADE MistralQR~\footnote{\spladis Fusion being an ensembling of two approaches, we report the maximum execution time of the two (as they could be executed in parallel).}. Note that retrieval is not optimized, and libraries such as PISA could reduce retrieval of SPLADE models below 100 ms on CPU, as shown in recent papers~\cite{10.1145/3477495.3531833/efficient,10.1145/3539618.3591941/pruning,twostepsplade}. This would further increase the efficiency gap between rewrite-based and distillation-based approaches. We also focus here on inference efficiency, as training efficiency does not directly affect the user, and would involve training LLMs. Those results provide a first answer to \ref{rq4} on the efficiency of \spladis compared to existing methods.

\begin{figure}[t]
    \centering
    \includegraphics[width=0.9\linewidth]{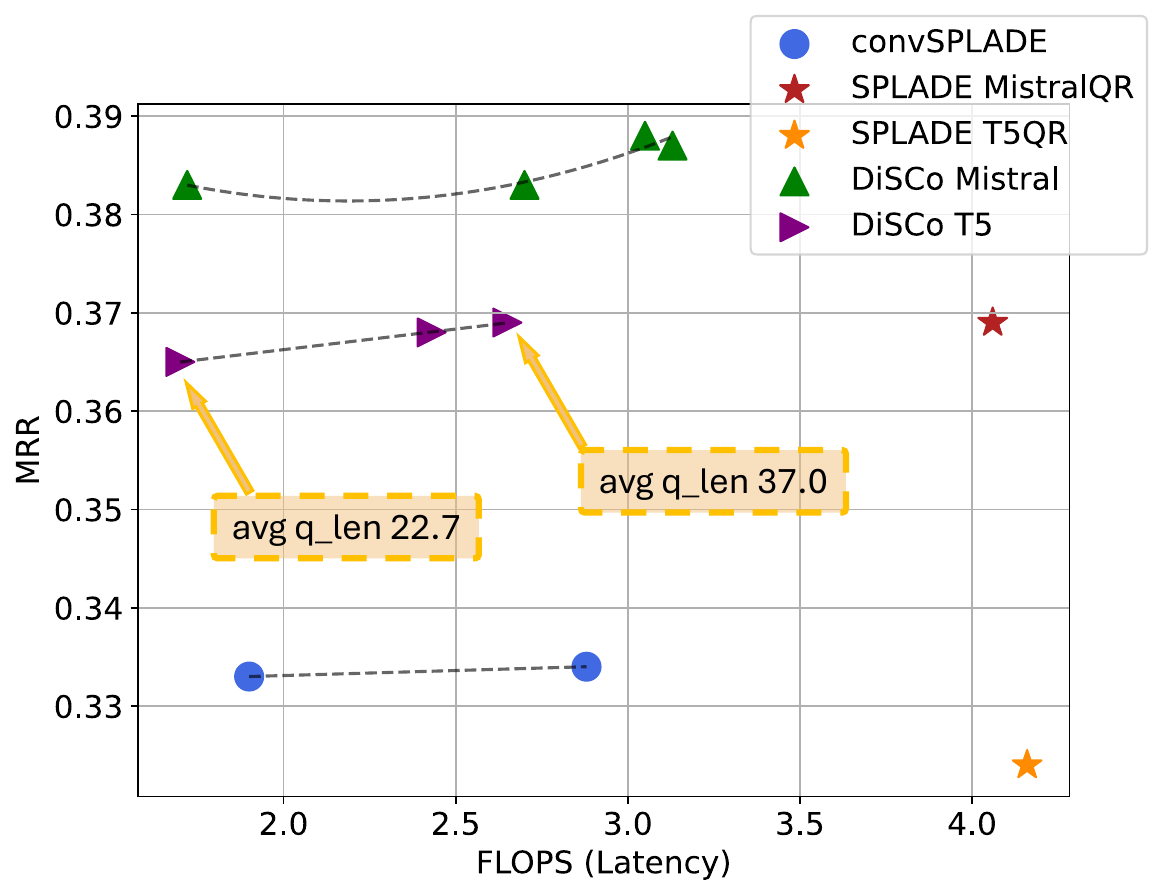}
    \caption{Effectiveness-efficiency trade-off on TopiOCQA. Sparser representations have a lower latency but also a lower MRR. \texttt{avg\  q\_len} is the average number of activated tokens in the conversation representations, as indicator of efficiency.}
    \label{fig:flops}
\end{figure}

\header{Controlling Sparsity.}
We then focus here on the sparsity of the sparse representations, as a measure of efficiency within the inverted index. Figure~\ref{fig:flops} plots the performance and FLOPs of several \spladis models when trained with different levels of sparsity. This is possible thanks to the regularization loss from the SPLADE architecture, controlled by the hyper-parameters $(\lambda_q, \lambda_d)$~\cite{spladev1}. From the Figure, we see for example that the same \spladis T5 model can be trained with different degrees of regularization, leading to different FLOPs and MRR. 
Furthermore, Figure~\ref{fig:flops} shows the FLOPs of the associated teacher models: SPLADE MistralQR and T5QR. We notice that our student \spladis models are more sparse compared to their teachers, as having only a constraint on similarities during training allows more freedom on the representations and their sparsity.
This answers part of \ref{rq4}, showing that \spladis can be trained with different sparsity levels.

\begin{figure}[t]
    \centering
    \includegraphics[width=\linewidth]{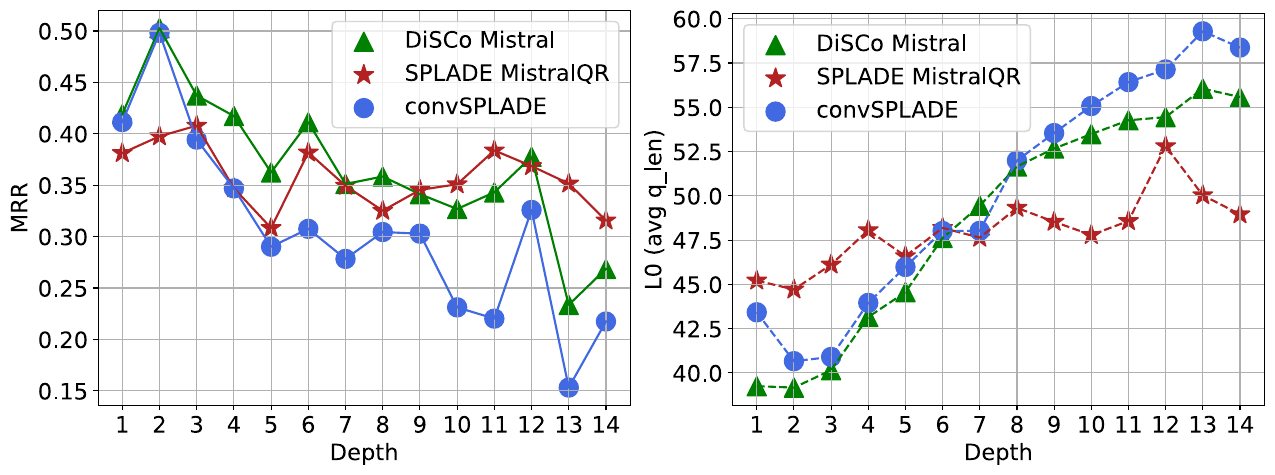}
    \caption{(Left) Performance with respect to depth on TopiOCQA. (Right) Sparsity of query representations with respect to depth of conversations.}
    \label{fig:depth}
\end{figure}

\header{Sparsity at different Conversation Depths.} To further address \ref{rq4}, we plot the sparsity and effectiveness of several baselines in Figure~\ref{fig:depth}. On the left part of the graph, we see the performance according to the depth of the conversation -- as the number of previous interactions between the user and the system -- and on the right the number of activated tokens in the representations. In particular, we see the difficulty for long conversations across models.  Comparing our \spladis with the convSPLADE baseline, we see that the main improvement comes from longer conversations. We also see that \spladis outperforms its teacher model, SPLADE MistralQR, on the first few turns of the conversation.
This is also a difficult task here as the TopiOCQA dataset has topic switches every few turns, making the context modeling task more complex.

Examining sparsity, we observe that the representations across all models remain sparse, with at most 60 non-zero dimensions even at deeper conversation levels ($>10$). At these depths, conversations become significantly longer, reaching up to 350 tokens. This suggests that the models effectively perform noise reduction as part of the context modeling task, activating only a subset of the tokens from the context. Now, comparing the models, as SPLADE MistralQR relies on Mistral rewrite, we see that the representation sparsity is consistent even for long conversations. For \spladis and convSPLADE, sparsity decreases with the turns of the conversation.

\section{Conclusion \& Future Work}

In this work, we propose \spladis, a novel distillation strategy in CS to distill query rewriting from LLMs. \spladis trains CS models with a relaxed distillation strategy that unifies context modeling and retrieval tasks. Our experiments on \ac{LSR} and \spladis models demonstrate that this training objective achieves important performance gains in both in-domain and out-of-domain retrieval. By distilling LLM rewrites, our method effectively learns from a single or several LLM teachers, outperforming the teachers, and reaching SoTA on several CS datasets. The proposed training strategy of \spladis also shows robustness to the quality of the teacher model when trained on weaker teacher models. We further examine the inference efficiency and sparsity of our approach after distillation.
Our findings emphasize the importance of aligning the rewriting and retrieval tasks in CS, with training objectives that unify both tasks.
As a future work, we plan to study different teachers, as any model producing a similarity score could be distilled, e.g., cross-encoder~\cite{hofstatter2020improving}, paired with stronger LLMs for rewrite.

\begin{acks}

This research was partly supported by the Swiss National Science Foundation (SNSF), under the project PACINO (Personality And Conversational INformatiOn Access), grant number 215742. This funding is a collaboration between the University of Amsterdam and the Università della Svizzera italiana.

\end{acks}

\bibliographystyle{ACM-Reference-Format}
\balance
\bibliography{sample-base}

\end{document}